\documentclass[showpacs,aps,prl,twocolumn,superscriptaddress]{revtex4-1}

\pdfoutput=1

\usepackage{amsmath,amsfonts,amssymb,bm,graphicx,hyperref,color}

\begin{document}

\title{{\small{Retrospective}}\\
Robert W. Zwanzig: Formulated nonequilibrium statistical mechanics}

\author{Hans C. Andersen} 
\email[Corresponding author, ]{hca@stanford.edu}
\affiliation{Department of Chemistry, Stanford University, Stanford CA 94305 USA }
\author{David Chandler}
\email[Corresponding author, ]{chandler@berkeley.edu}
\affiliation{Department of Chemistry, University of California, Berkeley, CA 94720, USA}

%\date{\today}

%\begin{abstract}
%BLAH BLAH
%\end{abstract}

\maketitle

Robert (`Bob') Zwanzig died on May 15 this year at the age of 86.  He had been a member of the National Academy of Sciences since 1972.  He is survived by Frances, his wife of 60 years, a daughter, Elizabeth Bennett, and a son, Carl.  Frances writes that Bob ``was a baker of bread and of made-from-scratch pizza, a maker of dry martinis and hearty soups, an ice dancer, and a walker along the Chesapeake and Ohio Canal.''  We remember him as the theorist who inspired us, sometimes criticized us, and more than anyone else, explained the foundations of nonequilibrium statistical mechanics.

Bob was born and raised in Brooklyn, New York.  He graduated from the Polytechnic Institute of Brooklyn in 1948 and then moved West, first to do experimental physical chemistry with Sidney Benson at USC, and then to do theory with John Kirkwood at Caltech.  Bob received his Ph.D. from Caltech in 1952, about a year after he had moved to Yale with Kirkwood in 1951.  That was when Bob was introduced to Frances.  She was a first-year graduate student at Yale.

Marshall Fixman, coming to Yale to do postdoctoral research with Kirkwood, first met Robert Zwanzig in early 1954.  He recalls hearing stories about him and other members of the Kirkwood group from the moment he arrived from MIT that January: 
``Berni Alder and Irwin Oppenheim were mentioned as great warriors ...
, [but] Bob Zwanzig was the current top gun, and it was suggested that murky thoughts should not be expressed in his presence.''  Zwanzig had already published an impressive work in 1951 with J. H. Irving on quantum hydrodynamics~\cite{Zwanzig1951}.  ``I was properly intimidated,'' Fixman says. 

Zwanzig returned from winter holiday in time to see Fixman's initiation in front of Kirkwood and Lars Onsager, among others.  ``At my seminar'' Fixman recalls, ``I talked about my PhD work 
... and seemingly replied to Onsager's objections with sufficient acumen that [Onsager] remained silent for the rest of the talk. Bob was inordinately pleased by this silence and decided on such flimsy evidence that I was worthy of toleration. 
... Bob and Frances, Bob Mazo and his wife Joan, and I had many good times hiking the hills around New Haven and swimming in Long Island Sound. They brightened [my] life .... 
Bob [Zwanzig] had great clarity of mind, as everyone appreciated, and also great warmth of heart.''

Fixman's remembrances hint at what was then the tradition in statistical mechanics --  rapid-fire discussion and persistent questioning 
modulated by a wholesome desire for understanding.  Throughout his career, Bob engaged in this type of activity with enthusiasm and vigor.  

That year, 1954, Bob published a paper introducing the idea that liquid-state properties could be estimated with thermodynamic perturbation theory~\cite{Zwanzig1954}. The 
importance of this work was not fully appreciated until computer simulation methods were applied more than a decade later to see
how accurate a theory based on this insight could be.

\begin{figure}
\begin{center}
\includegraphics[width=3.5in]{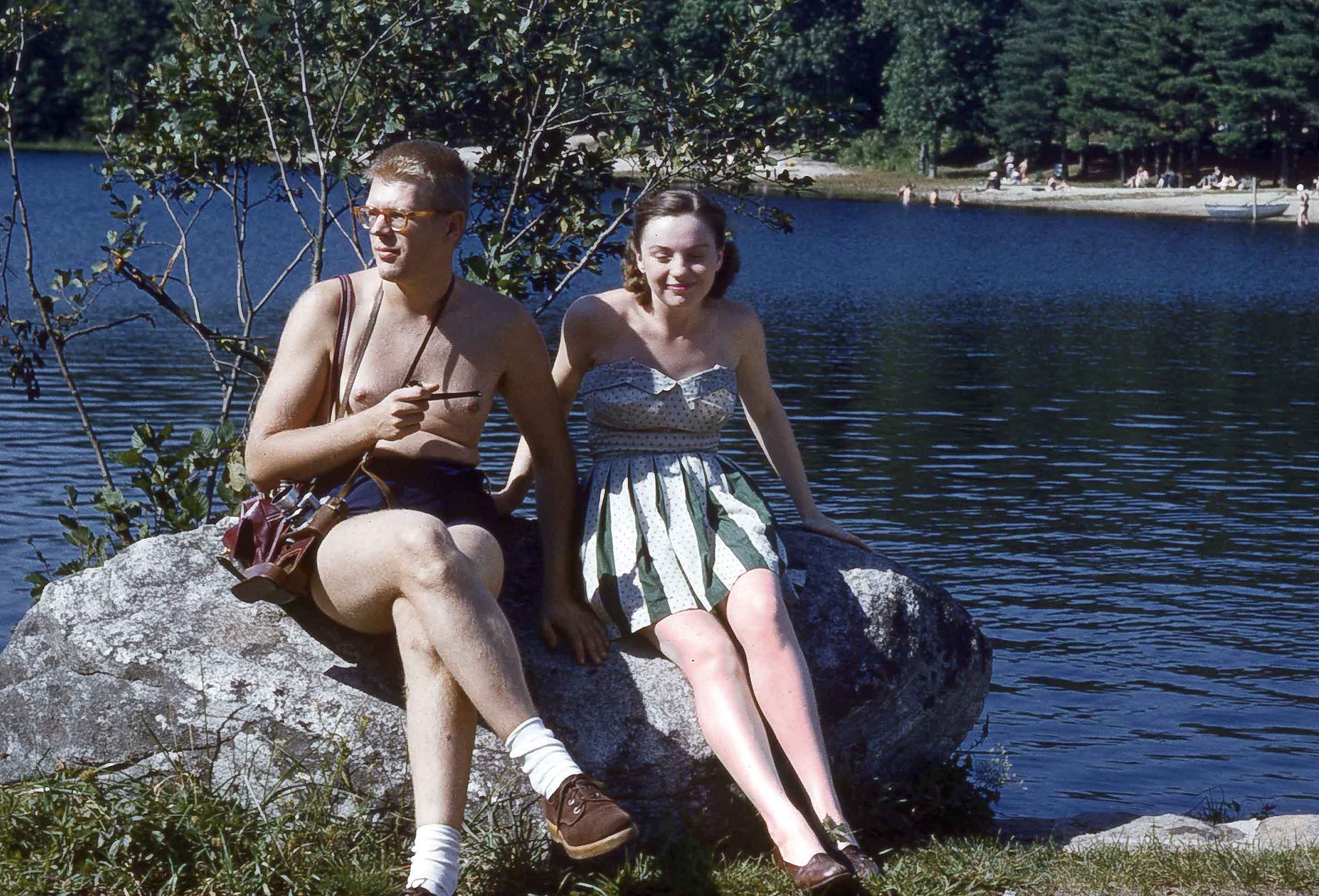}
\caption{ Robert and Frances (ne\'e Ryder) Zwanzig in the summer of 1954, during one of the outings recorded photographically by Marshall Fixman.  From 1983 to 1996, Frances Zwanzig served as Managing Editor of these \textit{Proceedings}.}
\label{RF54 }
\end{center}
\end{figure}

Zwanzig's greatest contributions came 
after he spent a few years on the faculty at Johns Hopkins and then took a position at the National Bureau of Standards.  
One of us (HCA) was a visiting graduate student in the Statistical Physics Section of the Bureau in 1963.  Zwanzig was in the Physical Chemistry Section, but he would frequently drop in to the Statistical Physics offices to talk with the postdocs about their research.  It was clear even to this graduate student that Zwanzig understood basic physics, chemistry, statistical mechanics, and mathematics much better than most scientists.  Moreover, Bob enjoyed talking with younger researchers about their interests and he shared his insights to aid them in their research. 
 While helping everyone else, Zwanzig was also formulating a perspective of nonequilibrium theory that would influence science for decades to come.

At the time, there was much confusion over how to treat irreversibility within a framework consistent with microscopic equations of motion, which are reversible.  There was qualitative appreciation of how Nature's arrow of time should emerge from 
probabilistic considerations.
With Onsager's insights and through the work of Ryogo Kubo, M. S. Green, and many others, microscopic expressions for transport coefficients and other material properties related to observed irreversible behavior had been derived, and these expressions contained time correlation functions of equilibrium systems.  But the derivations of these expressions were based on the assumption that the underlying microscopic behavior was irreversible.  It was not understood how these expressions
might be derived from reversible microscopic dynamics.

Ilya Prigogine and his co-workers offered explanations based upon summing infinite perturbation expansions of classical distribution functions, and L\'eon Van Hove did similarly for quantum density matrices.  These complicated developments seemed to require artful approximations often lacking clear physical interpretation, leading many to find their explanations less than compelling.  Zwanzig's ideas on a more transparent approach crystalized shortly after hearing Prigogine lecture on the topic~\cite{Zwanzig1992}.  

Zwanzig's essential idea was to use a projection operator method that permitted the derivation of a formally exact closed description of the dynamics of a chosen small set of variables of a material system that is initially out of equilibrium~\cite{Zwanzig1960}.  The equations contain well defined (but intractable) functions that depended on the details of the microscopic motions in the system. The derivation was based on reversible classical mechanics and did not assume anything about irreversibility.   He then showed how apparently irreversible behavior arose in certain classes of well defined physical situations \cite{Zwanzig1960,Zwanzig1961a}.  Special cases of this formalism led to the essential results of Prigogine and van Hove, as well as to a more complete understanding of the basis of Onsager's work and of the use of equilibrium time correlation functions to describe nonequilibrium irreversible behavior. 

His clarifying perspective on irreversibility, augmented by subsequent contributions from Hazime Mori and others, provided the basis for an incredible variety of work on the theory of irreversible processes in the 1960's and beyond.  Zwanzig himself participated in several of these advances, and he also helped the effort pedagogically by writing review articles that served as indispensable texts for our generation of statistical mechanicians~\cite{Zwanzig1961, Zwanzig1965}.  In 2001, Zwanzig published a textbook~\cite{Zwanzig2001} that organized much of what was accomplished during that decade and later.

His direct and elegant style of writing, seen in his basic papers on irreversibility and in so many others,
was also reflected in his lecturing.  The two of us recall several occasions 
when his lectures left us in awe.  Few matched his skill at elucidating the essential element of a difficult problem and then forging or adapting the mathematical tools required for its solution. 
His masterful style of lecturing, combined with his sometimes ascerbic style of public discussion, deterred some people from engaging him in private discussion.  Our experience over many years and many encounters was 
that Bob Zwanzig was consistently friendly and forthcoming when asked for scientific information or  insight with regard to a research problem.  Also, when asked for advice, he often offered wise counsel.

Zwanzig left the Bureau in 1966 to join the faculty of the University of Maryland.  He remained there for 22 productive years, contributing to a remarkable range of topics -- hydrodynamics, dielectrics, liquid crystals, liquid interfaces, polymer structure and dynamics, electronic energy transport in disordered materials and rate processes with dynamical disorder.  In 1988, Zwanzig moved to the Laboratory of Chemical Physics at the National Institutes of Health, where he made several important contributions to the fields of protein dynamics and protein folding, remaining active until a few years ago when poor health limited his ability to get to work.  
Colleagues from each of these institutions tell us that they will miss
him as a treasured friend and colleague -- a sometimes impatient man, but with magnificent perception and analytical skill, unquenchable curiosity and generosity.

\end{document}